\documentclass[prb,aps,twocolumn,showpacs, floatfix]{revtex4}

\usepackage[applemac]{inputenc}
\usepackage[T1]{fontenc} 
\usepackage{amsmath} 
\usepackage{lipsum}
\usepackage{amsfonts} 
\usepackage{amssymb, mathrsfs}
\usepackage{braket}
\usepackage{graphicx} 
\usepackage[usenames]{color} 

\usepackage{bbm}

\DeclareMathOperator\arcsinh{arcsinh}

\def\beq{\begin{equation}}
\def\eeq{\end{equation}}
\def\bsp{\begin{split}}
\def\esp{\end{split}}
\def\bea{\begin{eqnarray}}
\def\eea{\end{eqnarray}}
\def\ba{\begin{array}}
\def\ea{\end{array}}

\def\lb{\left(}
\def\rb{\right)}

\def\l.{\left.}
\def\r.{\right.}

\def\part{\partial}

\def\ket#1{\mid #1 {\cal{i}}}

 \numberwithin{equation}{section}

\begin{document}

\title{Coordinate (in)dependence and quantum inteference in quantum spin tunnelling}
\author{Solomon Akaraka Owerre}
\email{solomon.akaraka.owerre@umontreal.ca}
\author{M. B. Paranjape} 
\email{paranj@lps.umontreal.ca}
\affiliation{Groupe de physique des particules, D\'epartement de physique,
Universit\'e de Montr\'eal,
C.P. 6128, succursale centre-ville, Montr\'eal, 
Qu\'ebec, Canada, H3C 3J7 }

\begin{abstract}
\section*{Abstract}  
We show how to calculate the amplitude for quantum spin tunneling with a $z$ easy axis.  In this case, using the usual spherical polar angles to parametrize spin coherent states, typically the tunnelling  variable is the polar angle $\theta$, passing from 0 to $\pi$, while the azimuthal angle $\phi$ is imaginary and constant.  As the energy is constant along the tunnelling trajectory and normalized to vanish, and the Wess-Zumino term $S_{WZ} = is\int d\tau \thinspace\dot{\phi}(1-\cos\theta )$ also seems to vanish,  it is difficult to see how we  can obtain a non-zero contribution to the tunnelling amplitude.  Indeed, the instanton trajectory gives zero contribution.  We show how, the entire contribution to the tunnelling amplitude in fact comes from moving $\phi$ from zero to its complex value at the beginning of the tunnelling trajectory at $\theta=0$ and moving it back to zero at the end at $\theta=\pi$.  The Wess-Zumino term can also be expressed in a coordinate system independent manner as $S_{WZ} =is\int d\tau d\xi \left[\hat n\cdot (\partial_\tau\hat n\times\partial_\xi\hat n)\right]$, a format that does not appear to be well known or understood in the condensed matter literature.  We show in detail how this expression corresponds exactly to the more familiar, coordinate system dependent expression, and how to obtain the tunnelling amplitude directly from it.  

\end{abstract}


\maketitle


\section{Introduction} 
It appears that many authors have systematically avoided the analysis of ferromagnetic (anti-ferromagnetic) spin models with an  easy-axis chosen to align along the $z$ axis.  However the formulation of physical spin systems with an easy-axis is the simplest when the axis is taken along the $z$-direction.  For such systems which admit tunnelling, the corresponding coordinate in spin coherent state path integral is $\theta$, the instanton is in this variable, while $\phi$ is  always, necessarily, complex and is often just a constant.  Since the energy is necessarily also constant and then can be normalized to vanish along the instanton trajectory, the action for the instanton is determined entirely by the Wess-Zumino (WZ) \cite{wznw} term $S_{WZ} = is\int d\tau \thinspace\dot{\phi}(1-\cos\theta )$.  Then, it is hard to conceive of how the instanton can give a non-vanishing result for the tunnelling amplitude when $\phi$ is a constant.  This affords an explanation of why a $z$ easy axis coordinate system seems to be systematically avoided, one does not know how to do the calculation.   Aligning the coordinate system so that one has a  $x$ or $y$ easy-axis model, the instanton trajectory is in $\phi$ which then is real and  $\theta$  is necessarily complex and often constant.  The tunnelling amplitude comes from the calculation of the first term of the WZ term which then is real.  Furthermore it is obvious in this case, that the total derivative term remains imaginary and therefore generates any quantum phase interference, (such as, for example, the  suppression of energy splitting for half-odd integer spin demonstrated by D. Loss, D. P. Di Vincenzo, G. Grinstein, \cite{l} and by J. von Delft, C. Henley, \cite{k}).   This circumstance,  when a physical result is obtained on the basis of a total derivative term in the action, leaves one with an uneasy feeling as a total derivative term does not affect the equations of motion and could presumably be thrown away.    For the choice of coordinates with a $z$ easy-axis, the WZ action is either real (as $\phi$ is imaginary)  or zero (as $\dot\phi=0$), so it is not obvious where the quantum phase interference comes from.  In principle the choice of the coordinate system is made to simplify a problem,  however, the physical amplitude must not depend on this choice.  Here we show, by studying a specific model, that in order to recover the quantum interference when the easy axis is aligned with the $z$ axis, $\phi$ must be translated from $\phi=0$ to $\phi=\phi_R+ i\phi_I$ at first, before the instanton can mediate $\theta: 0\rightarrow\pi$, and then $\phi = \phi_R+ i\phi_I$  must be translated back to $\phi=0$ after the instanton  has occurred.  The contribution from the total derivative term in the WZ action for this round trip is exactly zero, but from the $\dot\phi\cos\theta$ we obtain the quantum interference.  We recover the results found in the references cited above.  We end with an exposition of the totally coordinate independent formulation.  We are able to solve the equations of motion for the instanton path, and recover the quantum phase interference by evaluating the Wess-Zumino term explictly.

In recent years single ferromagnetic spin systems have become a subject of interest due to the fact that they exhibit first- or second-order phase transition between quantum and classical regimes for the escape rate \cite{solo4, cl, solo6,chud1,solo5}. They also allow the possibility of   studying macroscopic quantum coherence (MQC) and macroscopic quantum tunnelling (MQT)\cite{c,gag,chud2}. The term ``macroscopic'' means that the system involves very large spin, therefore it can be described using a semi-classical approach. Both MQC and MQT usually involve two states separated by a barrier. In MQC,  tunnelling between neighbouring degenerate vacua is dominated by the instanton configuration with nonzero topological charge leading to an energy level splitting. Hence, the degeneracy is lifted, and the true ground state is then the coherent superposition of the two degenerate vacua. In MQT, however, tunnelling is dominated by the bounce configuration \cite{B} with zero topological charge leading to the decay of the metastable states.  The quantum tunnelling effect in spin systems occur both in ferromagnetic and anti-ferromagnetic spin system \cite{B,F,ams,chud,had}.  In ferromagnetic systems, the macroscopic variables satisfy the well known Landau-Lifshitz differential equation. 
 
The tunnelling rate (energy splitting) is often calculated semi-classically using the instanton method. This method  has been studied extensively in one dimension using the imaginary time path integral \cite{col}. For spin systems, however, the imaginary time path integral (the spin coherent-state path integral)  contains an additional phase that contributes to the transition amplitude. The phase appears because the overlap of two different coherent states is not unity. The Euclidean (imaginary time) action from this method contains two terms, one term is the spin (magnetic)  energy which is real. This  term is responsible for the energy barrier between two states, and the other term (the Wess-Zumino or Berry phase term) is imaginary, first-order in time derivative, and contains a topological (total derivative) term.  For a single spin model, the action is usually parametrized by two coordinates $\theta$ and $\phi$.  Since the action is complex, one of these variables has to be complex for the equation of motion to be consistent.  It was shown in a specific model \cite{l}, that when the real tunnelling coordinate is  $\phi$ ($\theta$ is complex), the topological term (which is imaginary)  causes destructive interference leading to the vanishing of tunnelling splitting when the total spin of a ferromagnet\cite{l,k} (or the excess spin of an antiferromagnet \cite{l,solo7,chud}) is a half-odd integer.  For systems with $z$-easy-axis\cite{fd,ow}, the real tunnelling coordinate is $\theta$, and $\phi$ is complex, and to obtain the quantum phase interference for half-odd integer spin is a bit subtle as in this case  the topological term is real or zero. In this  report, we show by a unique and elegant approach that the well known result, suppression of tunnelling for half-odd integer spin, can be recovered  for the $z$-easy-axis models.

 \section{Model and Method}
We will study the simple, single ferromagnetic spin Hamiltonian 
\bea
H=-K_zS_{z}^2 +K_yS_{y}^2, \quad K_z\gg K_y>0
\label{1}
\eea
using spin coherent state path integral.
The above Hamiltonian possesses an easy-axis in the $z$-direction and hard-axis along the $y$-direction, so we expect the real tunnelling variable to be $\theta$ which parametrizes the rotation in $z$-axis. The Hamiltonian has been studied in a magnetic field by many authors \cite{c,ams} . However, the quantum phase interference for this model has not been reported in any literature, we believe, due the subtlety involved in computing the action for the instanton.   

The Hamiltonian studied by  M Enz  and R Schilling  \cite{F}  
\bea
H=-AS_{x}^2 +BS_{z}^2,\quad (h=0)
\label{2.2}
\eea 
possesses an easy $x$-axis and a hard-axis along the  $z$- axis. This model in the conventional spherical parametrization $\bold{S} = (\sin\theta\cos\phi, \sin\theta\sin\phi, \cos\theta)$ is exactly our Hamiltonian Eq.\eqref{1} in the unconventional spherical parametrization $\bold{S} = (\sin\theta\sin\phi, \cos\theta, \sin\theta\cos\phi)$. In order to demonstrate our technique for investigating the quantum phase interference in $z$ easy-axis model, we will  stick to the conventional spherical parametrization. It was shown \cite{ga} that perturbation theory  in the $K_y$ term for integer spin leads to an energy splitting proportional to $(K_y)^s$ while for half-odd integer spin, the splitting vanishes in accordance with Kramers' theorem. We  will recover this result using spin coherent state path integral.

The transition amplitude in spin coherent state path integral is given by \cite{D} 
\begin{equation}
  \langle \theta_{f}, \phi_{f} |e^{-\beta H}|\theta_{i}, \phi_{i}\rangle= \int  \mathcal{D}\left[\cos\theta \right]
 \mathcal{D}\left[\phi \right] e^{-S_E/\hbar}
\label{10.2}
\end{equation}
where the Euclidean action is
\bea
S_E = \int d\tau \left[is \dot{\phi}(1-\cos\theta ) + E(\theta,\phi)\right]\label{act}
\eea
and the classical anisotropy energy Eq.\eqref{1} is
\begin{equation}
\begin{split}
E(\theta,\phi)=(K_z + K_y\sin^2\phi)\sin^2\theta .
\label{10.3}
\end{split}
\end{equation}
The classical degenerate ground states correspond to $\phi=0$, $\theta = 0,\pi$, that is the spin is pointing in the north or south pole of the two-sphere. The classical equations of motion obtain by varying the action with respect to $\theta$ and $\phi$ respectively are
\bea
  is\dot{\phi} \sin\theta &=& -\frac{\partial E\lb\theta, \phi\rb}{\partial \theta}
\label{14}\\
is\dot{\theta} \sin\theta &=& \frac{\partial E\lb\theta, \phi\rb}{\partial \phi}
\label{ca}
\eea
It is evident from these two equations, because of the explicit $i$,  that one variable has to be imaginary in order for the equations to be consistent. The only appropriate choice is to take real $\theta$ and imaginary $\phi$, since the real tunnelling coordinate ($z$-easy-axis) is $\theta$. This comes out naturally from the conservation of energy, which follows by multiplying Eqn. \eqref{ca} with $\dot\phi$ and Eqn. \eqref{14} by $\dot\theta$ and subtracting the two:
\beq 
\frac{d E\lb\theta, \phi\rb}{d\tau}= 0 \quad  \text{i.e},\quad
E\lb\theta, \phi\rb = \text{const.} = 0
\eeq
Thus,
\beq
E(\theta,\phi)=(K_z + K_y\sin^2\phi)\sin^2\theta=0
\label{16}
\eeq
Since $\sin^2\theta\neq 0$, it follows that,
\begin{equation}
\sin\phi = \pm i\sqrt{\frac{K_z}{K_y}},
\label{8}
\end{equation}
Therefore $\phi$ is imaginary and constant. Let $\phi = \phi_R + i\phi_I$, then $\sin\phi = \sin\phi_R\cosh\phi_I + i\cos\phi_R\sinh\phi_I$. We must take $\phi_R = n\pi$ as the RHS of \eqref{8} is imaginary. Hence
\begin{equation}
 (-1)^n\sinh\phi_I =\pm \sqrt{\frac{K_z}{K_y}},
\label{88}
\end{equation}
There are four solutions of this equation: $n=0$, $\phi = i\phi_I $ and $n=1$,$\phi = \pi-i\phi_I $ for the positive sign and $n=0$, $\phi = -i\phi_I $ and $n=1$, $\phi = \pi+i\phi_I $ for the negative sign. Taking into account that $K_z\gg K_y$, we have $\phi_I= \arcsinh\lb\sqrt{\frac{K_z}{K_y}}\rb\approx \frac{1}{2}\ln\lb\frac{4K_z}{K_y}\rb$. The classical equation of motion \eqref{ca} simplifies to
\bea
is\frac{\dot{\theta}}{\sin\theta} = K_y\sin2\phi= iK_y\sinh2\phi_I 
\eea
 The solution is easily found as 
\bea
\thinspace \theta\lb \tau\rb =  2 \arctan [ \exp(\omega (\tau-\tau_0))], \label{sth}
\eea
where $\omega=  {\frac{K_y}{s}}\sinh2\phi_I$.
This corresponds to the tunnelling of the state $\ket{\uparrow }$ from $\theta\lb \tau\rb=0$ at $\tau= -\infty$ to the state $\ket{\downarrow }$, $\theta\lb \tau\rb=\pi$ at $\tau = \infty$. 
The two solutions $\phi = i\phi_I$ and $\phi = \pi+i\phi_I$ in the upper half plane correspond to the instanton, ($\dot\theta>0$) while the solutions $\phi = -i\phi_I$ and $\phi = \pi-i\phi_I $ in the lower half plane correspond to the anti-instanton, ($\dot\theta<0$).

Since the energy, $E(\theta,\phi)$ in the action Eqn. \eqref{act} always remains zero along this trajectory the action for this path is determined only by the Wess-Zumino term which is given by
\bea
S_E=S_{WZ}= is\int_{-\infty}^{\infty}d\tau \thinspace\dot{\phi}(1-\cos\theta )
\label{aka}
\eea
We reiterate, if we had the real instanton trajectory in $\phi$, as would be the case for $x$ or $y$ easy-axis,  which interpolates between $\phi\lb \tau\rb=0$ at $\tau= -\infty$ and $\phi\lb \tau\rb=\pi$ at $\tau= \infty$, it is obvious that the total derivative term is imaginary then one can easily derive the quantum phase interference effect for which half-odd integer is suppressed \cite{l,k}. In the present analysis the instanton is not in $\phi$ but in $\theta$,  so care must be taken when computing the action.
Naively, one can use the fact that $\phi$ is constant and hence $\dot{\phi}=0$ which gives $S_{WZ}=0$. 


This fails to give a non-vanishing action. The problem can be rescued by using the technique that we recently employed\cite{solo}.   A non-vanishing action can only be obtained by taking into account that $\phi$ must be translated from $\phi=0$  to $\phi=n\pi+i\phi_I$ before the instanton can occur and then back to $\phi=0$ after the instanton has occurred. In the present problem, we have two solutions for $\phi$, i.e $\phi = i\phi_I$ and $\phi = \pi+i\phi_I$ corresponding to two instanton paths, call them $I$ and $II$. The full  action is then
\bea
S_E^I&=&is\int_0^{\pi+i\phi_I} \hskip-.8cm  d\phi(1 -\cos\theta)|_{\theta=0} +is\int^0_{\pi+i\phi_I}  \hskip-.7cm  d\phi (1 -\cos\theta)|_{\theta=\pi}\nonumber \\
&=&-2\pi is +2s\phi_I
\eea
and
\bea
S_E^{II}&=&is\int_0^{i\phi_I}d\phi(1 -\cos\theta)|_{\theta=0}  \\&+&is\int^0_{i\phi_I}  d\phi (1 -\cos\theta)|_{\theta=\pi}=2s\phi_I\nonumber
\eea
where it is clear that the total derivative term contributes nothing as the two contributions cancel in the round trip, while the $d\phi\cos\theta$ gives all the answer, since $\cos\theta=1$ before the instanton has occurred, while $\cos\theta=-1$ after.   
The amplitude for the transition from $\theta =0$ to $\theta= \pi$ can be calculated by summing over a sequence of one instanton followed by an anti-instanton with an odd total number of instantons and anti-instantons \cite{col}, but we must add the two exponentials of the actions $S_E^I$ and $S_E^{II}$ for both instanton and anti-instanton, we get that the expression for the amplitude is given by
\bea
\braket{\pi|e^{-\beta \hat{H}}|0}=\sinh\lb 2\kappa\beta(1+\cos(2\pi s))e^{-2s\phi_I}\rb
\label{trans}
\eea
where $\kappa$ is the ratio of the square root of the determinant of the operator governing the second order fluctuations, without the zero mode.
The energy splitting can be read off from this expression
\bea
\Delta E = 2\kappa(1+\cos(2\pi s))e^{-2s\phi_I} 
\eea
For half-odd integer spin the splitting vanishes while for integer spin we have
\bea
\Delta E=4\kappa\lb\frac{K_y}{4K_z}\rb^s
\eea
which agrees with the result found by perturbation theory \cite{ga}.

\section{Coordinate independent formalism}
In the coordinate independent formalism, the spin is represented by a unit vector $\hat n (\tau)$ but no parametrization of the unit vector is assumed.  Then the action for the Hamiltonian in Eqn.\eqref{1} can be written as
\bea
S_E=\int d\tau {\cal L}_E=\int d\tau \left[ -K_z(\hat n\cdot\hat z)^2+K_y(\hat n\cdot\hat y)^2\right]\nonumber\\
+is\int d\tau d\xi \left[\hat n\cdot (\partial_\tau\hat n\times\partial_\xi\hat n)\right].\label{actci}
\eea
The first term is the anisotropy energy while the second term is the Wess-Zumino term written in the its native, coordinate independent form.  The Wess-Zumino term is integrated over a two manifold whose boundary is physical, Euclidean time $\tau$.  Thus the configuration in $\tau$ is extended into a second dimension with coordinate $\xi$.  The equations of motion arise from variation with respect to $\hat n$.  However, $\hat n$ is a unit vector, hence its variation is not arbitrary, indeed, $\hat n\cdot \delta\hat n=0$.  Thus to obtain the equations of motion, we vary $\hat n$ as if it is not constrained, but then we must project onto the transverse  part of the variation:
\beq
\delta_{\hat n}S_E=0\,\,\Rightarrow \,\, \int d\tau (\delta_{\hat n}{\cal L}_E)\cdot\delta{\hat n}=0\,\,\Rightarrow \,\, \hat n\times(\delta_{\hat n}{\cal L}_E)=0
\eeq
Taking the cross product of the resulting equation one more time with $\hat n$ does no harm, and this process yields the equations of motion
\beq
is\partial_\tau\hat n-2K_z(\hat n\cdot\hat z)(\hat n\times\hat z) +2K_y  (\hat n\cdot\hat y)(\hat n\times\hat y)=0 \label{eomnh}           
\eeq
Taking the cross product of the equation with $\partial_\tau\hat n$, the first term vanishes as the vectors are parallel yielding
\beq
-2K_z(\hat n\cdot\hat z)\partial_\tau\hat n\times(\hat n\times\hat z)+2K_y(\hat n\cdot\hat y)\partial_\tau\hat n\times(\hat n\times\hat y)=0.
\eeq
Simplifying the triple vector product and using $\partial_\tau\hat n\cdot\hat n=0$, and taking the scalar product of the subsequent equation with $\hat n$ gives
\beq
\partial_\tau\left(-K_z(\hat n\cdot\hat z)^2+K_y(\hat n\cdot\hat y)^2\right)=0
\eeq
which is the conservation of energy.  From this equation and that $\hat n$ is a unit vector we find
\bea
\hat n\cdot\hat y&=&\pm\sqrt{\frac{K_z}{K_y}((\hat n\cdot\hat z)^2-1)}=\pm i\sqrt{\frac{K_z}{K_y}(1-(\hat n\cdot\hat z)^2)}\nonumber\\
\hat n\cdot\hat x&=&\pm\sqrt{\frac{K_y+K_z}{K_y}(1-(\hat n\cdot\hat z)^2)}
\label{hxhy}
\eea
where the $\pm$ signs are not correlated.
Then
\beq
\frac{\hat n\cdot\hat y}{\hat n\cdot\hat x}=\pm i\sqrt{\frac{K_z}{K_y+K_z}}\equiv \tan\phi
\label{phi}
\eeq
hence we recover the result immediately that  $\phi$ is an complex constant, just as before.  $\phi$ is of course just the usual azimuthal angle of the spherical polar coordinate system. Taking the scalar product of Eqn. \eqref{eomnh} yields
\beq
is\partial_\tau(\hat n\cdot\hat z) +2K_y(\hat n\cdot\hat y)(\hat n\cdot\hat x)=0
\eeq
and replacing from Eqn. \eqref{hxhy} gives
\beq
is\partial_\tau(\hat n\cdot\hat z) \pm i2\sqrt{K_z(K_y+K_z)}(1-(\hat n\cdot\hat z)^2)=0
\eeq
Notice that the $i$'s neatly cancel leaving a trivial, real differential equation for $\hat n\cdot\hat z$, which we can write as
\beq
\frac{\partial_\tau(\hat n\cdot\hat z)}{1-(\hat n\cdot\hat z)}+\frac{\partial_\tau(\hat n\cdot\hat z)}{1+(\hat n\cdot\hat z)}=\pm \frac{4}{s}\sqrt{K_z(K_y+K_z)}.
\eeq
This integrates as
\beq
\ln\frac{1+(\hat n\cdot\hat z)}{1-(\hat n\cdot\hat z)}=\pm \frac{4}{s}\sqrt{K_z(K_y+K_z)}(\tau-\tau_0).
\eeq
Exponentiating and solving for $\hat n\cdot\hat z$ gives
\beq
\hat n\cdot\hat z=\pm\tanh \lb \frac{2}{s}\sqrt{K_z(K_y+K_z)}(\tau-\tau_0)\rb
\eeq
which is exactly the same as the solution found for $\theta$  in Eqn. \eqref{sth}. The instanton (upper sign ) interpolates from $n_z=1$ to $n_z=-1$ as $\tau\rightarrow\pm\infty$.

Thus it is important to know that the equations of motion can be solved without recourse to a specific choice for the coordinates. We will now evaluate the tunnelling amplitude and the quantum interference directly in terms of the coordinate independent variables. Since the energy remains constant along the instanton trajectory, the action is determined entirely from the WZ term
\bea
S_{WZ} =is\int d\tau \int_{0}^{1}d\xi \left[\hat n\cdot (\partial_\tau\hat n\times\partial_\xi\hat n)\right]
\label{3.13}
\eea
The integration over $\xi$ can be done explicitly by writing the unit vector as
\bea
\hat n(\tau,\xi)= f(\tau,\xi)n_z(\tau)\hat z + g(\tau,\xi)[n_x(\tau)\hat x+n_y(\tau)\hat y]
\label{3.14}
\eea
with the boundary conditions $\hat n\lb\tau, \xi=0\rb=\hat n(\tau)$ and $\hat n\lb\tau,\xi=1\rb=\hat z$. Using the expression in Eq.\eqref{3.14} and the condition that $\hat n\cdot \hat n =1$ one obtains
\bea
g^2=\frac{1-f^2n_z^2}{1-n_z^2}
\label{3.15}
\eea
 These functions obey the boundary conditions
 \begin{align}
& f(\tau,\xi=0)=1, f(\tau,\xi=1)=\frac{1}{n_z(\tau)}, \nonumber\\& g(\tau,\xi=0)=1, g(\tau,\xi=1)=0 \end{align}

The integrand of Eq.\eqref{3.13} can now be written in terms of the functions defined in Eq.\eqref{3.14}. After a long calculation we obtain 
\begin{align}
\hat n\cdot (\partial_\tau\hat n\times\partial_\xi\hat n)&= (g^2f^{\prime}-fg g^{\prime})(n_x\dot{n}_y-n_y\dot{n}_x)\label{3.17}\nonumber\\& = \frac{n_zf^{\prime}}{1-n_z}(n_x\dot{n}_y-n_y\dot{n}_x)
\end{align}
where $f^{\prime}\equiv \partial_\xi f$, $\dot{n}_{x,y}\equiv \partial_\tau{n}_{x,y} $. The second equality follows from Eq.\eqref{3.15}.  Replacing Eq.\eqref{3.17} into the WZ term, the $\xi$ integration in Eqn. \eqref{3.13} can be done explicitly which yields
\bea
S_{WZ} =is\int d\tau \frac{(n_x\dot{n}_y-n_y\dot{n}_x
)}{1+n_z}
\label{3.13a}
\eea
This expression defines the WZ term in the coordinate independent form as a function of time alone. We can always make recourse to any specific coordinates, taking the $z$ easy-axis system, with the spherical parameterization
one recovers the usual form of the WZ term use in condensed matter physics i.e Eq. \eqref{aka}. Multiplying the top and the bottom of the integrand in Eq.\eqref{3.13a}  by $(1-n_z)$, the resulting integrand simplifies to
\begin{align}
S_{WZ} &=is\int \frac{d(n_y/n_x)}{1+(n_y/n_x)^2}(1-n_z) \label{3.13b}\nonumber\\& = is\int d[\arctan(n_y/n_x)](1-n_z)\nonumber\\& = is\int d\phi(1-\cos\theta)
\end{align}
This shows how to obtain the usual spherical polar coordinate system dependent expression for the WZ term from the coordinate independent one.   

It is noted from Eq.\eqref{phi} that $\phi$ has to be imaginary. In order to recover the quantum phase interference in the coordinate independent formalism, $\phi$ must to translated from the initial point say $\phi=0$ to the final point $\phi=n\pi +i\phi_I$, $n=0,1$ before and after the instanton occurs \cite{solo}. The two contributions to the action  from these paths are given by
\bea
S_{WZ}^I&=&is\int_0^{\pi+i\phi_I} \hskip-.8cm  d\phi (1-n_z)|_{n_z=1} +is\int^0_{\pi+i\phi_I}  \hskip-.7cm  d\phi(1-n_z)|_{n_z=-1}\nonumber \\
&=&-2\pi is +2s\phi_I
\eea
and
\bea
S_{WZ}^{II}&=&is\int_0^{i\phi_I}d\phi (1-n_z)|_{n_z=1}\\&+&is\int^0_{i\phi_I} d\phi (1-n_z)|_{n_z=-1}=2s\phi_I\nonumber
\eea
which are the exact expressions as before. Then the previous evaluation quantum interference goes through unchanged. 

\section{Conclusion}
We have investigated a biaxial ferromagnetic spin model with $z$-easy axis as an exemplary system where tunnelling occurs along the $z$ axis.   For this model, we found that the real instanton trajectory is in $\theta$ while $\phi$ is imaginary.  Since the action for the trajectory is completely determined by the Wess-Zumino term, which in the present problem is either real or zero, it is not clear where the suppression of tunneling for half-odd integer spin comes from. We showed that for this model there are four complex solutions for $\phi$ of which two correspond to an instanton and the other two correspond to an anti-instanton, therefore there are two instanton and anti-instanton paths. The quantum phase interference is obtained by translating $\phi$ from zero to these complex solutions and then back to zero, the exponentials of the two actions add and give rise to a factor of $(1+\cos(2\pi s))$ in the energy splitting, which obviously vanishes for half-odd integer. We also explicitly solved for the instanton and its corresponding action in the coordinate independent formalism showing how the result is completely independent of the coordinate system. The quantum phase interference was recovered exactly as before.

\section{  Acknowledgments } 
We thank  NSERC of Canada and the Direction des relations internationales de l'Universit\'e de Montr\'eal for financial support. We also thank Yassine Hassouni of the D\'epartment de physique, Facult des Seciences, Universit\'e Mohammed V Agdal, Rabat, Morocco, for hospitality, where this work was done.  



\begin{thebibliography}{99}
%
 \bibitem{wznw}
J. Wess, B. Zumino, Phys. Lett. B {\bf 37}:95,1971; S.P. Novikov,  Usp.Mat.Nauk,  37N5:3-49,1982; E. Witten, Nucl. Phys., B{\bf 160}:57,1979. 
 \bibitem{l}
D. Loss, D. P. Di Vincenzo, G. Grinstein, \prl {\bf 69}, 3232 (1992)
\bibitem{k}
J. von Delft, C. Henley, \prl {\bf 69}, 3236(1992)

\bibitem{cl} 
E.M. Chudnovsky and D.A. Garanin, \prl {\bf 79}, 4469 (1997).
\bibitem{solo4}
Gwang-Hee Kim, \prb {\bf 59},  11847, (1999); J. Appl. Phys. 86, 1062 (1999) 
\bibitem{solo5}
J.-Q. Liang, H. J. W. M\"{u}ller-Kirsten, Y.-B. Zhang, Jian-Ge Zhou, F. Zimmerschied   and F.-C. Pu  , \prb {\bf 57}, 529 (1998) 
\bibitem{solo6}
Y.-B. Zhang, J.-Q. Liang, H.J.W. M\"{u}ller-Kirsten, S.-P. Kou, z.-B. Wang and F.-C. Pu \prb {\bf 60}, 12886 (1999)
 \bibitem{chud1}
 D. A. Garanin, X. Mart\`inez Hidalgo, and E. M. Chudnovsky  \prb {\bf  57}, 13639 (1998)
\bibitem{c} 
E.M. Chudnovsky and L. Gunther, \prl {\bf 60}, 661(1988); \prb {\bf 37}, 9455 (1988).
 \bibitem{chud2}
 E. M. Chudnovsky and J. Tejada, Macroscopic Quantum Tunnelling of Magnetic Moment, Cambridge university press (1998) 
  \bibitem{gag}
E. M.  Chudnovsky, and B.  Barbara, and P. C .E  Stamp Int. J. Mod. Phys. B{\bf 6},1355 (1992)
\bibitem{B} 
E.M. Chudnovsky and B. Barbara, Phys. lett. A {\bf 145}, 205 (1990).  

 \bibitem{ams}
Anupam Garg and Gwang-Hee Kim, \prb {\bf 45}, 12921 (1990)
 \bibitem{had}
A. Garg, EuroPhys. Lett. {\bf 22}, 205 (1993)
\bibitem{F} 
 Enz M and Schilling R (1986) J. Phys. C: Solid State Phys. 19 L711-5;1765-70
 \bibitem{chud}
 Eugene M. Chudnovsky Journal of Magnetism and Magnetic Material {\bf  140},1821 (1995)

\bibitem{col}
 Sidney Coleman, Aspects of symmetry, Cambridge university press (1985)
 
\bibitem{solo7}
Yi-Hang Nie, Yan-Hong Jin, J-Q Liang, H J W M\"{u}ller-Kirsten, D K Park, F-C Pu, J. Phys.: Condens. Matter {\bf 12} (2000) L87-L91
\bibitem{fd}
 Florian Meier and Daniel Loss \prl{\bf 86}, 5373 (2001)
\bibitem{ow}
 O. Waldmann, C. Dobe, H. U. G\"{u}del, and H. Mutka   \prb {\bf 74}, 054429 (2006)
 \bibitem{ga}
D A Garanin (1991) J. Phys. A: Math. Gen. 24 L61

  \bibitem{D}
 John R. Klauder \prd{19}, 2349 (1978)
\bibitem{solo}
Solomon Akaraka Owerre and M.B. Paranjape, \prb {\bf 88}, 220403(R) (2013)
 

  
 

 \end{thebibliography}
\end{document}